\documentclass[preprint]{elsarticle}

\usepackage{booktabs, ctable, dcolumn}
\usepackage{epsfig, amsmath}
\usepackage{datetime}
\usepackage{graphicx}    
\usepackage{graphicx}
\usepackage{ulem}
 \usepackage{multirow}
 \usepackage{verbatim}
\newcommand{\be}{\begin{equation}}
\newcommand{\ee}{\end{equation}}
\newcommand{\bea}{\begin{eqnarray}}
\newcommand{\eea}{\end{eqnarray}}

\newcommand{\Rmnum}[1]{\expandafter\@slowromancap\romannumeral #1@}

\begin{document}

\begin{frontmatter}

\title{Van der Waals force assisted heat transfer for vacuum gap spacings}
\author[label1]{K. Sasihithlu \corref{cor1} }
\author[label1]{J. B. Pendry}
\author[label2]{R. V. Craster}

\address[label1]{The Blackett laboratory, Imperial College London, London SW7 2AZ, UK}

\address[label2]{Department of Mathematics, Imperial College London, London SW7 2AZ, UK}

\cortext[cor1]{k.sasihithlu@imperial.ac.uk}


\end{frontmatter}

\date{\ddmmyyyydate \today}

\section{Abstract}

Phonons (collective atomic vibrations in solids) are more effective in transporting heat than photons. This is the reason why the conduction mode of heat transport in nonmetals (mediated by phonons) is dominant compared to the radiation mode of heat transport (mediated by photons). However, since phonons are unable to traverse a vacuum gap (unlike photons) it is commonly believed that two bodies separated by a gap cannot exchange heat via phonons. Recently,  a mechanism was proposed \cite{john2016phonon} by which  phonons can transport heat across a vacuum gap - through Van der Waals interaction between two bodies with gap less than wavelength of light.  Such heat transfer mechanisms are highly relevant for heating (and cooling) of nanostructures; the heating of the flying heads in magnetic storage disks is a case in point. Here, the theoretical derivation for modeling phonon transmission is revisited and extended to the case of two bodies made of different materials separated by a vacuum gap. Magnitudes of phonon transmission, and hence the heat transfer, for commonly used materials in the micro and nano-electromechanical  industry are calculated and compared with the calculation of conduction heat transfer through air for small gaps.

\section{Introduction}

The analysis of interaction between two objects when placed close together (smaller than wavelength of light) has led to the observation of new and interesting phenomena like near-field radiative heat transfer where the radiative transfer between the objects exceeds Planck's blackbody limit by several orders of magnitude.  The theoretical description and experimental confirmation of this phenomenon has given rise to new applications like thermal radiation scanning tunneling microscopy \cite{de2006thermal}, near-field thermophotovoltaics \cite{laroche2006near}, and non-contact radiative cooling \cite{guha2012near}. Recently it has been recognized that radiative (mediated by photons) exchange cannot be the only mode of near-field heat exchange between two closely spaced bodies. In an experimental study \cite{altfeder2010vacuum} to understand the thermal coupling between a scanning tunneling microscope tip and a gold substrate, for a spacing of $\approx$ 0.3 nm vacuum gap, a heat flux  six orders of magnitude larger than predictions of near-field radiation theory was observed. In this paper we describe the possibility of an additional channel of heat transfer due to phonon transmission across vacuum gap. Since it can be shown that the number of {\it propagating} modes available for heat transfer via phonons is approximately $(c/c_l)^2$ times greater than that available for photons, where $c$ [$c_l$] is the speed of light [sound] in vacuum [solid], and since  the velocity of sound in most solids is of the order of $10^3$ m$\text{s}^{-1}$,  
 if a mechanism can exist where these phonons can transmit across the vacuum gap, it can be expected to be a significant source of heat transfer between two objects separated by vacuum.  Such a mode of transmission of phonons across a vacuum gap is far from obvious since phonons, being the quanta of lattice vibrations, require a material medium to propagate. The basic premise of this work is that when two bodies are brought in close proximity to each other in vacuum, the van der Waals interaction between them can
 act as a conduit for the phonons to propagate across the vacuum gap.  While this Van der Waals interaction is weak compared to the inter-atomic bonds that exist in a solid,  the sheer number of propagating modes available for heat exchange via phonons compared to photons could make this mechanism a significant source of heat transfer. Along with the propagating modes,  modes which are trapped on the surface of the objects can also contribute to heat transfer, as is observed in the photon exchange process.

There have been a few recent attempts to estimate the heat transport via phonon transport across vacuum gap. 
 Sellan et. al. \cite{sellan2012phonon} observed that the heat transfer via phonons can be four orders of magnitude higher than that via near-field radiative transfer for a gap of 1 $\AA$ between two silicon surfaces. However this lattice-dynamics based simulation took into account only the interactions of the surface atoms via their electron clouds and not the relatively long-range Van der Waals interaction between objects for the phonon transmission. Prunnila and Meltaus \cite{prunnila2010acoustic} predicted significant phonon transmission when closely spaced mediums are made up of piezoelectric material. This effect is due to the electric field induced by the phonons in the material acting as a conduit for the phonon transmission.  The fact that Van der Waals forces can act also as a conduit for phonon exchange, and hence heat transfer, across the vacuum gap has been recognized in two recent works \cite{ezzahri2014vacuum,budaev2015heat}. However due to inconsistencies in these works  we believe that the correct picture of phonon transmission across the vacuum gap has not yet been captured.  Ezzhahri et. al.,  observed that for highly doped silicon it is possible for phonon mediated heat transfer to dominate the radiative transfer. But they consider only ballistic transport of phonons across the gap and do not elucidate on the role of different modes of elastic waves that can exist in a solid like compressional, transverse and surface waves.  Moreover, the phonon transmission is calculated by modeling the interaction of atoms across the vacuum gap to be spring-like, which is strictly valid only at very small gaps $\approx 1 \, \AA$  where repulsive forces due to overlapping electron clouds start to become significant. Budaev and Bogy do not assume spring-like interaction, but consider only compressional elastic waves in their study and ignore other possible modes; no numerical data has been presented regarding how dominant the heat exchange via phonons is vis-\`a-vis radiative transfer at small gaps. A more recent work by Chiloyan et. al. \cite{chiloyan2015transition} follows the method adopted by Ezzhahri et. al. of modeling the Coulombic interaction of atoms across the vacuum gap with a spring-like behavior to calculate phonon transmission, and show that for subnanometer gaps the contribution from low frequency acoustic phonons can be significant.  A common drawback in these works is the non-consideration of sinusoidal variation of the surface topology of the solids due to the presence of phonons.

\begin{figure}
  \includegraphics[scale=0.3]{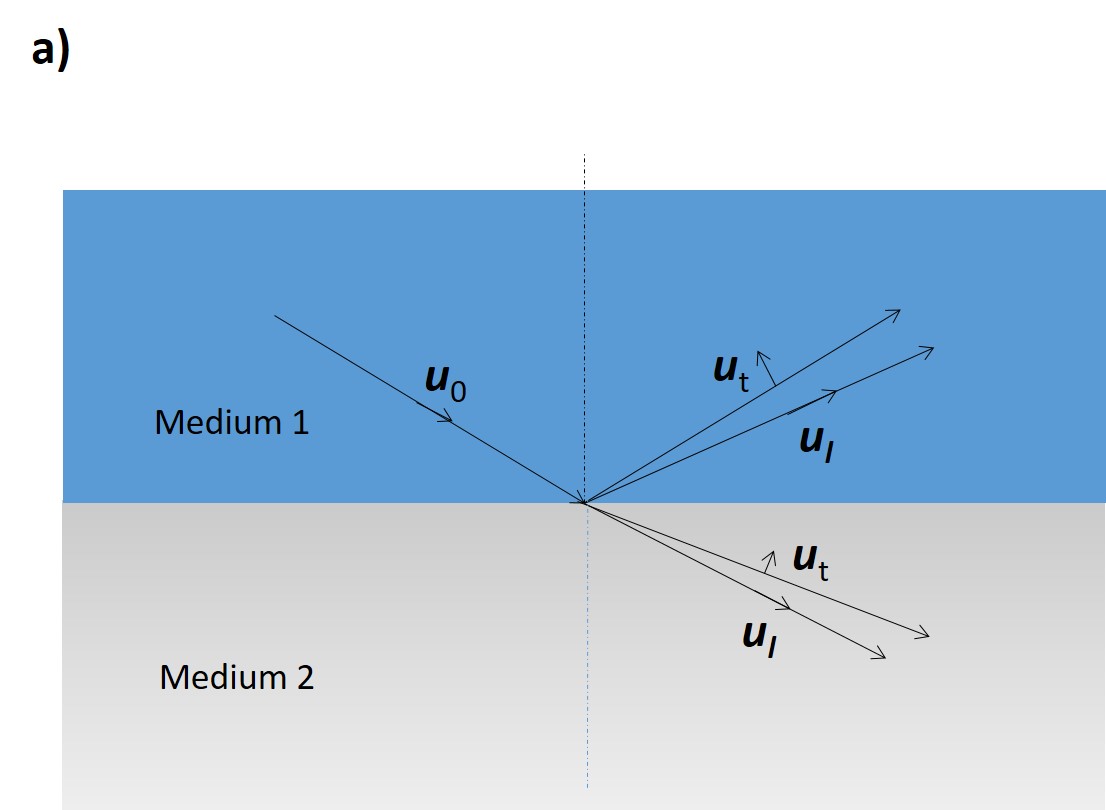}
  \includegraphics[scale=0.3]{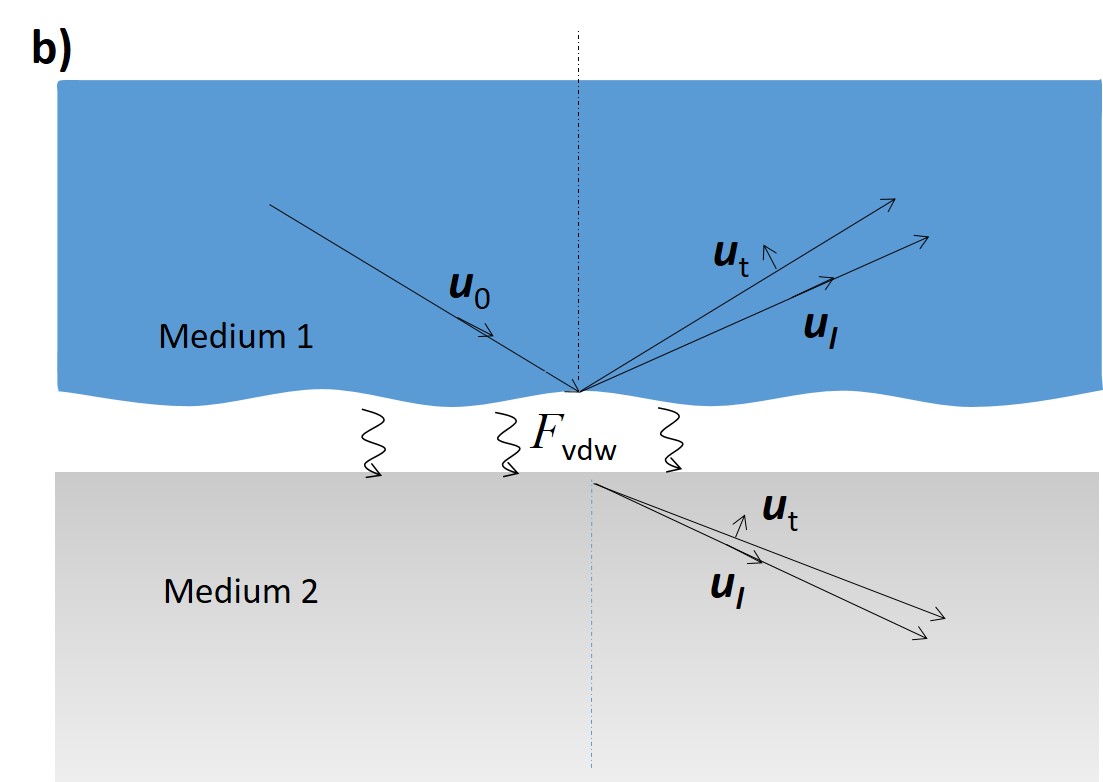}
  \caption{{\small {\bf (a)} Acoustic mismatch method, developed to analyse the thermal boundary resistance at an interface between two media, considers the elastic waves to undergo specular reflection and transmission at the interface. Here, $\boldsymbol u_l$ and $\boldsymbol{ u_t}$ are unit vectors along the directions of the displacement of the atoms and indicate the longitudinal and transverse polarization modes respectively. {\bf (b)}  The configuration of two closely spaced planar surfaces which  interact via Van der Waals forces ($F_{\text{vdw}}$).  Due to the presence of phonons, the surface in medium 1 is not flat but has a sinusoidal variation with frequency depending on the parallel component of the wave-vector of the phonon. } }
  \label{fig}
\end{figure}

A model which takes into consideration these effects was described in Ref. \cite{john2016phonon}, where expressions were derived for the transmission coefficients of phonons when two objects made of the same material are brought in close proximity to each other. The aim of the current paper is to describe in more detail the model put forward in \cite{john2016phonon}, extend this work to provide expressions for transmission coefficients when objects of different materials are separated by small spacings, provide numerical estimates of heat transfer between bodies made of materials commonly used in nano and microelectromechanical industry where such effects are expected to be important, and to understand where this mode of heat transfer becomes important in real-life situations by comparing this with conduction heat transfer when the gap is filled with air. To model the phonon transmission across a vacuum gap  we adopt the theory of estimating the thermal boundary resistance (or the Kaptiza resistance) at an interface between two media and extend it to our configuration of two bodies separated by a vacuum gap. A study of heat transfer across an interface between two media such as that shown in Fig. 1(a) is of critical importance in systems where high heat dissipation is crucial (for example: in semiconductor devices). Hence, such a system has been studied in extensive detail and models have been developed to estimate heat transfer across such an interface; one such model is the acoustic mismatch method \cite{swartz1989thermal}. The methodology developed is as follows: phonons are modeled as propagating elastic waves and the two possible polarizations modes, longitudinal  and transverse, are  assumed to undergo specular reflection and transmission at the interface. Since each polarization mode can give rise to a combination of modes at the interface there are four unknowns to be solved for (two reflection and two transmission coefficients) and these are determined using the stress and displacement boundary conditions at the interface. The transmission coefficients thus obtained determine the amount of heat that gets transmitted across the interface. Importantly, the heat transfer coefficient across an interface predicted from this method has been experimentally verified to hold true for well prepared interfaces and for a wide range of temperatures \cite{swartz1989thermal, anderson1981kapitza}.
While extending this procedure to calculate the heat transfer for our configuration shown in Fig. 1(b) of closely spaced bodies separated by vacuum and at different temperatures, we employ boundary conditions that are different from that adopted in the acoustic mismatch method (due to the presence of vacuum gap and the normal Van der Waals force at the surface of the solids).   An important and novel aspect of our work is to take into account the modulation of the surface of the solids due to the presence of the collective atomic vibrations (phonons)  and the resulting effect on the interaction between the surfaces. The surface of the hot body with phonons (Medium 1 in Fig. 1(b)) will have a sinusoidal variation with spatial frequency equal to the parallel component of the wave-vector of the phonons. 
Assuming for simplicity that the Van der Waals force satisfies additive principle, this sinusoidal variation on the surface will be analytically shown  to impose an exponential limit on the number of wave-vectors which contribute to interaction between the surfaces.  Prior works have failed to recognize this effect so that the heat transfer contributions from the higher wave-vector modes would be over-estimated. Since we ignore phonon-phonon interactions and work in the continuum limit  the method proposed here is valid for temperatures that are small compared to the Debye temperature where only low frequency acoustic phonons are activated.

The paper is arranged as follows: In Sec. 3 a general expression for the van der Waals force is derived for an arbitrary sinusoidal displacement of the surface. In Sec. 4, an expression for the amplitude of the sinusoidal displacement at the surface of a planar object is derived as a function of incident angle and the polarization of the incident elastic wave, which, when used in the result from Sec. 3 gives the value of the force acting on the surface of the planar object. This force is then related to the stress in the media and expressions for transmission coefficients of the phonons as a function of incident angle are  derived from the boundary conditions. In Sec. 5, expression for the heat transfer is derived and numerical values of heat transfer for objects made of materials commonly used in the micro electromechanical industry are compared.

\section{van der Waals stresses on the surface}

\begin{figure}[h]
\includegraphics[scale=0.35]{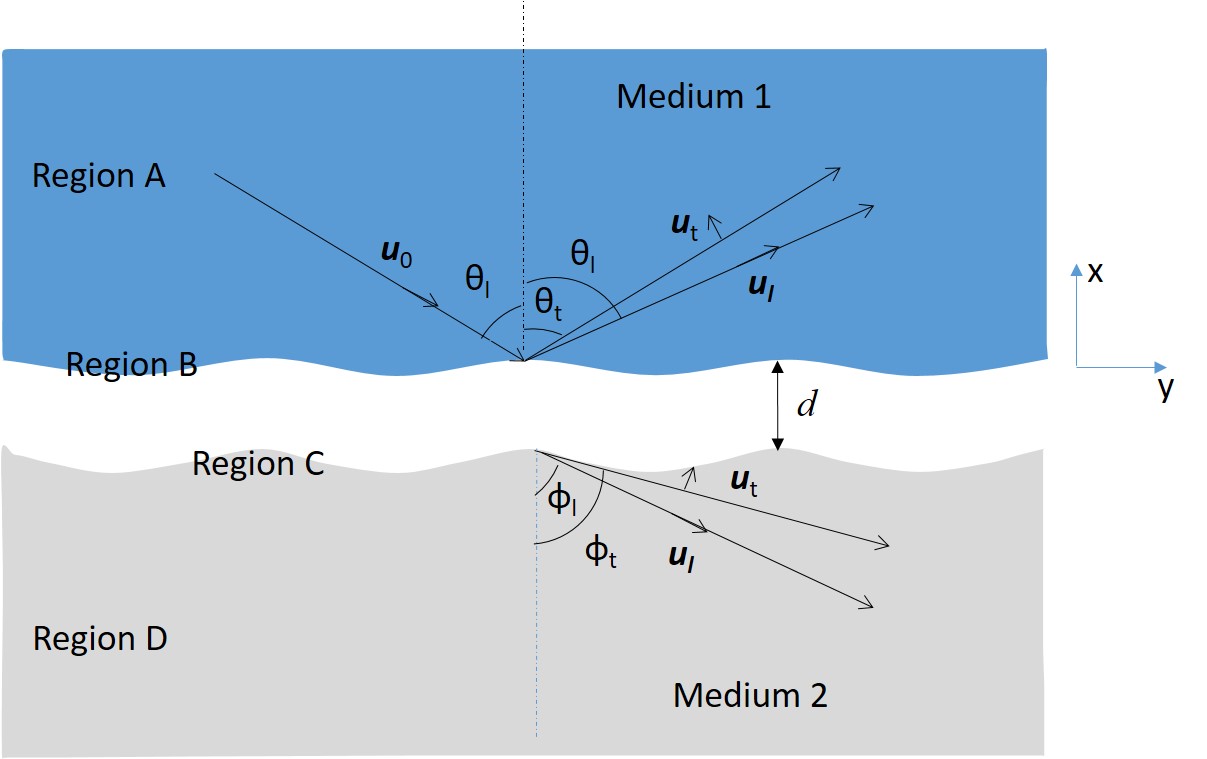}\protect\caption{In our configuration the two media denoted by Medium 1 and Medium 2 are separated by a distance $d$. 
Region B(C) comprises the section of medium 1(2) where the x-coordinate varies from $x=d(0)-\delta/2$ to $x=d(0)+\delta/2$ where $\delta$ is the amplitude of the sinusoidal displacement on the surface}
\label{fig2}
\end{figure}

We consider one of the bodies, say medium 1, to be held at a finite temperature $T$ and the other at 0 K so that the effect of phonons in lattice deformation is prevalent  only in medium 1.  To find the time-varying van der Waals force acting on the surface of medium 2, we consider a test displacement on its surface and calculate the total van der Waals potential of the resulting configuration. The total potential between the two media can be divided into bulk-bulk (interaction between region A and region D in Fig. \ref{fig2}), bulk-surface (interaction between region A and region C, and between region B and region D in Fig. \ref{fig2}), and surface-surface (interaction between region B and region C) contributions. Since the bulk-bulk interaction potential is time-invariant it does not contribute to phonon transmission, and only the bulk-surface  and the surface-surface  interaction will be considered separately below.

Consider first the region B-region C interaction potential $\langle\phi\rangle_{B-C}$. If $\rho_n$ is the number density of molecules in the two media, then the number of atoms in an elementary volume $dV$ of material is $\rho dV$. Let $\beta$ be the dispersion constant accounting
for interaction between two atoms in the opposite mediums. Assuming for simplicity that the potential satisfies additivity principle,  $\langle\phi\rangle_{B-C}$ can be got by accounting for the pair-wise interaction between atoms in the two sinusoidal displacements on the surface of the two media and, in the limit $\delta_{1x}$, $\delta_{2x} \ll d$,  is given by:

\begin{equation}
\langle\phi\rangle_{B-C}=\frac{1}{2}\rho_n^{2}\beta \int_{{\bf r_{||}}}\int_{\mathbf{R_{||}}}\frac{\delta_{1x}\delta_{2x}\cos({\bf k_{||}.{\bf r_{||}}})\cos({\bf k_{||}.R_{||}}))}{[|\mathbf{r}_{||}-\mathbf{R}_{||}|^{2}+d^{2})]^{3}}d^{2}{\bf r_{||}}d^{2}\mathbf{R_{||}}
\label{eq:langle}
\end{equation}
where, the factor $\frac{1}{2}$ is introduced noting that integration would involve counting the pair-wise interaction twice; the displacement at the surface of the two media are given by: $u_x^{(1)} =\delta_{1x} \cos({\bf k_{||}.{\bf r_{||}}})$; and $u_x^{(2)} =\delta_{2x} \cos({\bf k_{||}.{\bf R_{||}}})$.
The dispersion constant $\beta$ can be related to the experimentally determined and catalogued Hamaker's constant H using \cite{israel91}:

\[
\frac{1}{2}\,\rho_n^2 \beta=\frac{\text{H}}{\pi^{2}}
\]
Since the integral
\[
\text{Re \ensuremath{\int_{{\bf r}_{||}}}\ensuremath{\frac{e^{i{\bf k_{||}}\cdot{\bf r_{||}}}d^{2}{\bf r_{||}}}{\left(|{\bf r}_{||}-{\bf R}_{||}|^{2}+d^{2}\right)^{3}}}}
\]
evaluates to
\[
2\pi\,\,\, \text{Re}\,\, e^{i{\bf k_{||}}\cdot{\bf R_{||}}}\int_{\rho=0}^{\infty}\frac{J_{0}(k_{||}\rho)\rho d\rho}{(\rho^{2}+d^{2})^{3}} = \frac{\pi}{4}\,\,\frac{k_{||}^{2}}{d^{2}}K_{2}(k_{||}d)\,\,\cos\left({\bf k_{||}}\cdot{\bf R_{||}}\right)
\]
where $\rho =|{\bf r}_{||}-{\bf R}_{||}|$, and $K_{2}(x)$ is the 2nd order Bessel function of the second kind, 
 Eq. \ref{eq:langle} thus reduces to the form:
\begin{equation}
\langle\phi\rangle_{B-C}=\frac{\text{H}\,\, L^2}{8\pi}K_{2}(k_{||}d)\dfrac{k_{||}^{2}}{d^{2}} \delta_{1x}\delta_{2x} \label{eq:potSingleIntegral}
\end{equation}
 where $L^{2}$ is the area of interaction between the two surfaces.
Now consider the region A-region C interaction potential $\langle\phi\rangle_{A-C}$ given by the expression:
\begin{equation}
\langle\phi\rangle_{A-C} = \frac{1}{2} \rho_n^{2}\beta \int \frac{d^3 {\bf R}_1\,d^3 {\bf R}_2}{|{\bf R}_1 - {\bf R}_2|^6}
\end{equation}
where ${\bf R}_1$ and ${\bf R}_2$ are the position vectors of two infinitesimally small volumes in region A and region C. By integrating over the bulk in region A and the sinusoidal distortion in region C, expression for $\langle\phi\rangle_{A-C}$ can be shown to reduce, in the second order approximation, to the form:
\begin{equation}
\langle\phi\rangle_{A-C} = -\frac{ \text{H}\,\, L^2 \delta_{1x}^2}{4 \pi d^4}
\end{equation}
Similarly, the expression for the interaction potential between region B and region D $\langle\phi\rangle_{B-D}$ reduces to the form:
\begin{equation}
\langle\phi\rangle_{B-D} = -\frac{ \text{H}\,\, L^2 \delta_{2x}^2}{4 \pi d^4}
\end{equation}

The total interaction potential between the two bodies is given by: $\langle\phi\rangle= \langle\phi\rangle_{B-C}+ \langle\phi\rangle_{A-C}+\langle\phi\rangle_{B-D}$. The normal van der Waals force acting on the surface (per unit area of interaction) of medium 1 and medium 2, $F_1$ and $F_2$, can be got from the interaction potential as:
\begin{align}
\label{eq:force}
F_1 &= - L^{-2}\frac{\partial \langle\phi\rangle }{\partial \delta_{1x}} =-\frac{\text{H}}{4\pi}K_{2}(k_{||}d)\dfrac{k_{||}^{2}}{d^{2}} \delta_{2x} + \frac{ \text{H}\,\,\delta_{1x}}{2 \pi d^4}; \\
 F_2 &= - L^{-2} \frac{\partial \langle\phi\rangle }{\partial \delta_{2x}} = -\frac{\text{H}}{4\pi}K_{2}(k_{||}d)\dfrac{k_{||}^{2}}{d^{2}} \delta_{1x} + \frac{ \text{H}\,\, \delta_{2x}}{2 \pi d^4}
\end{align}
The presence of the factor $K_{2}(k_{||}d)$ imposes a limit on the contribution of large wave-vector modes to the interaction between the two surfaces, particularly for large gaps. This can be seen from the asymptotic approximation for the function $K_{2}(k_{||}d)$ \cite{abramowitz}, where, in the limit $k_{||}d \gg 1$, $K_{2}(k_{||}d) \approx \sqrt{\pi/(2k_{||}d)} e^{-k_{||}d} $. Due to this exponential cut-off in the large wave-vector modes, the heat transfer falls off at a higher rate than that calculated previously in Ref. \cite{ezzahri2014vacuum, chiloyan2015transition} where this effect has been ignored.
As observed in Eq. \ref{eq:force} the expressions for the normal van der Waals forces  is linear in the displacement amplitudes $\delta_{1x}$ and $\delta_{2x}$.  The shear-force between the two surfaces can be obtained by considering the sinusoidal distortion to have a phase difference $\Phi$ and differentiating with respect to $\Phi$. A similar method has been adopted in Ref. \cite{chen2002demonstration}.  This gives the expression for the shear force to be quadratic in the displacement amplitudes, and since we are in the limit of small amplitudes, it can be neglected in relation to the normal van der Waals force.


%
%
%

\section{Transmission coefficient for the phonons}
As mentioned in Sec. 1, to find the heat transmitted to the second medium due to the van der Waals interaction between the surfaces, we take the continuum limit where phonons can be modeled as elastic waves. To relate the expressions of the Van der Waals force found in Sec. 1 to the stress in the medium (taken to be isotropic) we look at the displacements in the medium in the presence of elastic waves. While it is possible to have three different polarizations modes - one longitudinal and two transverse (polarized in the horizontal and vertical planes, also referred to as SH and SV waves respectively), since the transverse mode polarized in the horizontal plane (SH mode) does not result in out-of-plane deformation it does not contribute to phonon transmission, and we only consider the  SV mode henceforth. Consider first an incident planar longitudinal wave of unit amplitude in medium 1 as shown in Fig. 1(b). The equation of the incident wave is of the form  $ {\bf u_0}=  {\bf a_0} e^{-i k_{lx} x}   e^{-i k_{y} y}$ where ${\bf a_0}$ is a unit vector in the direction of propagation of the incident wave and can thus be expressed in terms of the incident angle $\theta_l$ as ${\bf a_0} =  -\cos \theta_l {\bf x}   -\sin \theta_l {\bf y} $, $k_{lx}$ and $k_{y}$ are the components of the wave-vector perpendicular and parallel to the surface respectively.  The displacement vector of the two reflected waves can be written as:  $ {\bf u_l}^{(1)} = R_l {\bf a_l}^{(1)} e^{+i k_{lx} x}   e^{-i k_{y} y}$  and $ {\bf u_t}^{(1)} =  R_t ({\bf z} \times {\bf a_t}^{(1)}) e^{+i k_{lx} x}   e^{-i k_{y} y}$, where $R_l$ ($R_t$) is the reflection coefficient of the longitudinal (transverse) component; and ${\bf a_l}^{(1)}$ (${\bf a_t}^{(1)}$) is a unit vector in the direction of propagation of the reflected longitudinal (transverse) wave.
%
Thus the components of displacement in medium 1 along the $x$ and $y$ directions, $u_x^{(1)}$ and $u_y^{(1)}$, are given by:
\begin{align}
\label{eq:dispmed1}
u_x^{(1)}  &= \big[- \cos \theta_l  e^{-i k_{lx} x}    + R_l \cos \theta_l e^{i k_{lx} x}    + R_t \sin \theta_t e^{i k_{tx} x}   \big] e^{-i k_{y} y}\\
u_y^{(1)}  &= \big[(- \sin \theta_l)  e^{-i k_{lx} x}   + R_l (- \sin \theta_l) e^{i k_{lx} x}    + R_t \cos \theta_t e^{i k_{tx} x}   \big] e^{-i k_{y} y}
\end{align}

Since the parallel component of the wave-vector has to be conserved across the interface, the angles $\theta_l$ and $\theta_t$ are related by: $\sin \theta_l c_t = \sin \theta_t c_l$. The stresses, $\sigma_{xx}^{(1)}$ and $\sigma_{xy}^{(1)}$ in medium 1 can be obtained from the normal and shear strains, $u_{xx}^{(1)}$, $u_{yy}^{(1)}$, and $u_{xy}^{(1)}$ \cite{landau1986} as:
\begin{align}
\sigma_{xx}^{(1)} &=  2 \rho c_t^2 u_{xx}^{(1)}   + \rho (c_l^2 - 2 c_t^2) u_{ll}^{(1)} \label{eq:stressnormal} \\
\sigma_{xy}^{(1)} &= 2 \rho c_t^2 u_{xy}^{(1)} \label{eq:stressshear}
\end{align}
where $u_{ll} = u_{xx} + u_{yy}$, $\rho$ is the density, $c_l$ and $c_t$ are the longitudinal and transverse velocities of sound in the isotropic material. This gives the stress at the surface ($x=0$), which should be equal to the van der Waals stress derived in Sec. 3, to be:
\begin{align}
\sigma_{xx}^{(1)} &=  \Big[ i \rho c_t^2  \left(    \cos 2 \theta_l k_l     +   R_l  \cos 2 \theta_l  k_l     + R_t  \sin 2 \theta_t  k_t   \right)  + i \rho (c_l^2 -  c_t^2)  k_l (1 + R_l) \Big] e^{-i k_{y} y} \label{eq:stressnormalcoeffmed11} \\
\sigma_{xy}^{(1)} &=  i\rho c_t^2 \Big[  (1 - R_l) k_l  \sin 2 \theta_l     +  R_t k_t \cos 2 \theta_t      \Big] e^{-i k_{y} y} \label{eq:stressnormalcoeffmed12}
\end{align}

In the second medium, for ease of analysis we initially assume the material to be the same as in medium 1,  before extending to the case of different materials.
The angles $\phi_l$ and $\phi_t$ in Fig. \ref{fig2} are thus equivalent to the angles $\theta_l$ and $\theta_t$.
  The displacement vector in the second medium due to the presence of transmitted longitudinal and transverse waves can be written as:
    $  {\bf u}^{(2)} = T_{l} {\bf a}_l^{(2)} e^{- i k_{lx} x} e^{- i k_{ly} y} +  T_{t} ({\bf z} \times {\bf a}_t^{(2)})  e^{- i k_{lx} x} e^{- i k_{y} y}$ giving us the components of the displacement vector along the $x$ and $y$ axes to be:
\begin{align}
    u_x^{(2)} &= \big[ T_{l} (-\cos \theta_{l}) e^{- i k_{lx} x}  + T_{t}  \sin \theta_{t} e^{- i k_{tx} x} \big] e^{- i k_{y} y}  \label{eq:dispmed21}\\
    u_y^{(2)} &= \big[ T_{l} (-\sin \theta_{l}) e^{- i k_{lx} x}   + T_{t}  (- \cos \theta_{t}^{(2)}) e^{- i k_{tx} x} \big] e^{- i k_{y} y} \label{eq:dispmed22}
     \end{align}

From Eq. \ref{eq:force},\ref{eq:dispmed1} and \ref{eq:dispmed21} we get the normal force acting on the surface (x=0) of medium 1, $F_1$, to be:
\begin{align}
F_1 = -\frac{\text{H}}{4\pi}K_{2}(k_{||}d)\dfrac{k_{||}^{2}}{d^{2}}\big[ -T_l \cos \theta_l +T_t \sin \theta_t \big] e^{-i k_{y} y}  +\frac{\text{H} }{2 \pi d^4}\big[- \cos \theta_l      + R_l \cos \theta_l    + R_t \sin \theta_t    \big] e^{-i k_{y} y}
\label{eq:F1coeff}
\end{align}

    and the expression for the normal force acting on the surface (x=0) of medium 2, $F_2$, is:
     \begin{align}
     F_2 = -\frac{\text{H}}{4\pi}K_{2}(k_{||}d)\dfrac{k_{||}^{2}}{d^{2}} \big[- \cos \theta_l      + R_l \cos \theta_l    + R_t \sin \theta_t    \big] e^{-i k_{y} y}  +\frac{\text{H} }{2 \pi d^4}\big[ -T_l \cos \theta_l +T_t \sin \theta_t \big] e^{-i k_{y} y}
     \label{eq:F2coeff}
     \end{align}
    From Eqs. \ref{eq:dispmed21} and \ref{eq:dispmed22} expressions for the compressive and shear stress at the surface of the second medium can be arrived as:
     \begin{align}
     \sigma_{xx}^{(2)} &=  \Big [\rho c_t^2  T_{l} i k_{l}   \cos 2 \theta_l   - i \rho c_t^2 k_{t} R_{t}  \sin 2\theta_{t}   + \rho (c_l^2 - c_t^2)  T_{l} i k_{l}  \Big] e^{- i k_{y} y} \label{eq:stressnormalcoeffmed21}\\    \sigma_{xy}^{(2)} &=  \rho c_t^2  \Big[ T_{l}  i k_{l}  \sin 2 \theta_{l}  + T_{t} i k_{t} \cos 2 \theta_{t}  \Big]e^{- i k_{y} y} \label{eq:stressnormalcoeffmed22}
     \end{align}

Equating the stresses at the surface to the van der Waals force gives us the four boundary conditions:
\begin{align}
& \sigma_{xy}^{(1)}   = 0\\
&  \sigma_{xx}^{(1)}  = F_1 \\
&\sigma_{xy}^{(2)} =0\\
&  \sigma_{xx}^{(2)} =   F_2
\end{align}
where, expressions for stresses $\sigma_{xx}^{(1)}$, $\sigma_{xy}^{(1)}$,  $\sigma_{xx}^{(2)}$ ,   $\sigma_{xy}^{(2)}$, and the forces $F_1$ and $F_2$ in terms of the reflection and transmission coefficients are substituted from Eqs. \ref{eq:stressnormalcoeffmed11}, \ref{eq:stressnormalcoeffmed12}, \ref{eq:stressnormalcoeffmed21},  \ref{eq:stressnormalcoeffmed22}, \ref{eq:F1coeff} and \ref{eq:F2coeff} respectively.
The four equations can be solved analytically to obtain the transmission coefficients. Analytical expressions for the transmission coefficients are:
\begin{align}
T_l^{(l)} = \frac{-2 Q_1 Q_4 }{(Q_1 + Q_2 + Q_3) (Q_1 + Q_2 - Q_3) +Q_4^2)};\,\,\,\,\ T_t^{(l)} = -T_l^{(l)} \frac{k_l \sin 2\theta_l}{k_t \cos 2\theta_t}
\label{eq:Tl}
\end{align}
where the superscript has been included to indicate the polarization of the incident wave (longitudinal, in this case).
A similar analysis for an incident transverse wave gives the transmission coefficients:
\begin{align}
T_l^{(t)} = \frac{(Q_5 - Q_2)Q_4 }{ (Q_1 + Q_2 + Q_3)(Q_1 - Q_3) + Q_4^2 };\,\,\,\,\ T_t^{(t)} = -T_l^{(t)} \frac{k_l \sin 2\theta_l}{k_t \cos 2\theta_t}
\label{eq:Tt}
\end{align}
where,
\begin{align}
Q_1 &= i \rho k_l c_t^2 \cos 2\theta_l+i \rho k_l(c_l^2 - c_t^2)\\
Q_2 &= i \rho k_l c_t^2 \frac{\sin 2 \theta_t  \sin 2 \theta_l}{\cos 2\theta_t}\\
Q_3 &= -\frac{\text{H}}{4\pi}K_{2}(k_l \sin \theta_l d)\dfrac{k_l^2 \sin^2 \theta_l}{d^{2}}  \left(\cos \theta_l + \frac{k_l \sin 2\theta_l \sin \theta_t}{k_t \cos 2 \theta_t}\right)\\
Q_4 &= \frac{\text{H} }{2 \pi d^4} \left(\cos \theta_l + \frac{k_l \sin 2\theta_l \sin \theta_t}{k_t \cos 2 \theta_t}\right)\\
Q_5 &= 2 i k_t \rho c_t^2 \sin 2\theta_t
\end{align}

These expressions are equivalent to that derived in Ref. \cite{john2016phonon}. This analysis can be extended to the case when you have different materials across the vacuum gap. The corresponding transmission coefficients are then given by:
\begin{align}
T_l^{(l)} = \frac{-2 Q_4 Q_1  }{(S_1 + S_2+S_3) (Q_1 + Q_2-Q_3)+ S_4 Q_4 };\,\,\,\,\ T_t^{(l)} = -T_l^{(l)} \frac{c_t' \sin 2\phi_l}{c_l' \cos 2\phi_t}
\label{eq:Tlchange}
\end{align}
\begin{align}
T_l^{(t)} = \frac{(Q_5 - Q_2)Q_4 }{ (S_1 + S_2 + S_3)(Q_1 - Q_3) + S_4 Q_4 };\,\,\,\,\ T_t^{(t)} = -T_l^{(t)} \frac{c_t' \sin 2\phi_l}{c_l' \cos 2\phi_t}
\label{eq:Ttchange}
\end{align}
where the additional factors $S_1$, $S_2$, $S_3$ and $S_4$ are given by:
\begin{align}
S_1 &= i \rho' k_l' c_t'^2 \cos 2\phi_l+i \rho' k_l'(c_l'^2 - c_t'^2)\\
S_2 &= i \rho' k_l' c_t'^2 \frac{\sin 2 \phi_t  \sin 2 \phi_l}{\cos 2\phi_t}\\
S_3 &= -\frac{\text{H}_{\text{eff}}}{4\pi}K_{2}(k_l' \sin \phi_l d)\dfrac{k_l'^2 \sin^2 \phi_l}{d^{2}}  \left(\cos \phi_l + \frac{k_l' \sin 2\phi_l \sin \phi_t}{k_t' \cos 2 \phi_t}\right)\\
S_4 &= \frac{\text{H}_{\text{eff}} }{2 \pi d^4} \left(\cos \phi_l + \frac{k_l' \sin 2\phi_l \sin \phi_t}{k_t' \cos 2 \phi_t}\right)
\end{align}
Here, $\rho', c_t'$ and $c_l'$ are the density, transverse velocity of sound, and longitudinal velocity of sound in medium 2 respectively. $\text{H}_\text{eff}$ is the effective Hamaker's constant for the two objects made of different materials, an estimation for which can be got from the Lifshitz formula \cite{israel91}:
\begin{equation}
H_{\text{eff}} = \frac{3}{2}k_B T \left(\frac{\varepsilon_1 -1}{\varepsilon_1 +1}\right) \left( \frac{\varepsilon_2 -1}{\varepsilon_2 +1}  \right) + \frac{3h}{4 \pi}\int_0^\infty \left(\frac{\varepsilon_1(i \nu) -1}{\varepsilon_1(i \nu) +1} \right) \left( \frac{\varepsilon_2(i \nu) -1}{\varepsilon_2(i \nu) +1}  \right) \,\, d\nu
\end{equation}
where, $\varepsilon_1$ and $\varepsilon_2$ are the static dielectric constants of medium 1 and medium 2 and $\varepsilon(i \nu)$ denotes the values of the dielectric functions at imaginary frequencies.

\section{Heat transfer}

The energy transmission coefficients can be obtained from the expressions for the transmission coefficients in Eqs \ref{eq:Tl} and \ref{eq:Tt} as:
  \begin{align}
\tau^{(i)}(\theta_i) = \frac{|T_l^{(i)}|^2 c_l' \cos \phi_l+|T_t^{(i)}|^2 c_t' \cos \phi_t}{c_i \cos \theta_i};\,\,\,\,\,\, i=l,t
  \end{align}
 In k-space since a volumetric element is given by:  $k^2 \sin \theta \,\, d \theta\,\, d \phi \,\, dk$, by populating these states with the Bose-Einstein distribution function we obtain the total number of phonons as a function of the incident angle $\theta$:
\begin{equation}
N(\theta)  = \frac{V}{8 \pi^2}  \int_{k =0}^{k_D } \frac{1 } {e^{\hbar c_j k/(\,\,k_B T)} - 1}  k^2 \sin \theta \,\, d \theta\,\,  dk
\label{eq:N}
\end{equation}
where $V$ is the total volume of the body. The largest possible wavevector $k_D$ in this model  is related to the lattice constant $a$ in the material with the relation: $4/3 \pi k_D^3 =(2 \pi/a)^3 $.  The net heat transfer from one body to the other is then obtained by considering the net flux of phonons into the second surface and  summing up over the two possible polarization modes, longitudinal and transverse \cite{pollack1969kapitza, little1959transport}:
%
 \begin{equation}
\begin{split}
 Q_{1 \rightarrow 2}(T) =   \frac{1}{8\pi^2 } \sum_{i=l,t}   \int_{\theta_i = 0}^{\pi/2} \int_{\omega =0}^{\omega_D } \frac{(\omega^2/c_i^3) \sin \theta_i } {\,e^{\hbar \omega/(k_B T)} - 1}  \tau^{(i)}( \theta_i) \,  \,\, c_i \cos \theta_i\,\, \times \hbar  \omega \,\, d\omega\,\,d \theta_i\,\,   
 \end{split}
 \label{eq:Q}
 \end{equation}

\begin{figure}[h]
\includegraphics[scale=0.45]{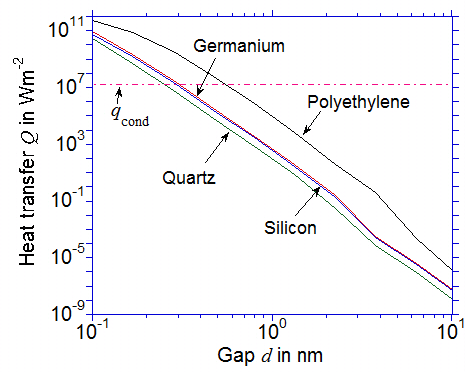}\protect\caption{Heat transfer calculated from Eq. \ref{eq:Q} between two semi-infinte planar surfaces made of the same material plotted as a function of gap between them. Values of material properties are tabulated in Table 1. The value of $q_\text{cond}$ is obtained from Eq. \ref{qcond}}
\label{fig3}
\end{figure}

In addition to the bulk longitudinal and transverse modes it is also possible for surface modes (Rayleigh waves) to contribute to the heat transfer. However, this has been shown in Ref. \cite{john2016phonon} to contribute negligibly to heat transfer and hence discussion on  this mode of heat transfer has been omitted here.
In Fig. \ref{fig3} we compare the heat transfer plotted as a function of gap between two bodies made of some commonly used materials in the nano and micro-electromechanical industry. Semiconductors like (silicon and germanium), metals (like gold and silver), polymers (like polyethylene and polystyrene),  ceramics (like quartz and silicon carbide) are some of the commonly used materials in this industry.  For example, quartz, due to its thermal stability, is a material-of-choice for making sensors; silicon and germanium serve as active substrates due to their dimensional stability to environmental conditions; polymers, due to their ease of processing and light weight, are widely used for device and machine components.
For the calculation of heat transfer, the temperature of one of the bodies is taken to be 300 K while the other is held at 0 K.

\begin{table}[h!]
\footnotesize
\begin{tabular}{|c|c|c|c|c|c|}
\hline
Material & H ($10^{-20}\,$ J) & $c_l$ (m/s) &  $c_t$ (m/s) & $\rho$ (g/cm$^3$)& $a$ (Angstrom)\\
\hline
Silicon & 25.5 & 8433 & 5843 & 2.30 & 5.431\\
Germanium & 30.0 & 5410 & 3350& 5.32& 5.68\\
Quartz & 7.93 & 5759 & 3765 & 2.66& 4.91\\
Polyethylene & 10.0 & 2100 & 850  & 0.9 & 2.55\\
\hline
\end{tabular}
\caption{Physical parameters including the Hamaker's constant H, the longitudinal velocity of sound $c_l$, the transverse velocity of sound $c_t$, the density $\rho$, and the lattice constant $a$, for a few materials used in nano and micro-electromechanical industry.}
\end{table}

To get an idea when the heat transfer due to phonon transmission can become significant in real-life situations, we compare our calculations with the heat transfer due to conduction in air for nanometer spacings, such as that between the writing head and the disk in magnetic storage devices. For gaps less than the mean free path of air molecules ($\approx 0.065 \mu$m at atmospheric pressure and room temperature \cite{karniadakis2006microflows}) heat transfer due to conduction, $q_\text{cond}$, does not follow the Fourier law from continuum theory but a separate law which gives precedence to boundary scattering can be derived from kinetic theory of gases \cite{han2005size, devienne1965low}:
\begin{equation}
q_\text{cond}=\frac{4 p k_B a^2 (T_1 - T_2)}{\sqrt{4 \pi k_B M } (\sqrt{T_1'}+\sqrt{T_2'})(2a-a^2)}
\label{qcond}
\end{equation}
where, $p$ is the pressure of gas, $k_B$ the Boltzmann constant, $a$ is the thermal accommodation coefficient which accounts for the interaction between the gas molecule and the two surfaces at temperatures $T_1=300$ K and $T_2=0 $ K, $M$ is the mass of a gas molecule ($\approx 4.8\times 10^{-26}$ kg for air) and the functions $T_1'$ and $T_2'$ are given by:
$T_1' = (a T_1 + a (1-a)T_2)/(2a - a^2)$,  $T_2' = (a T_2 + a (1-a)T_1)/(2a - a^2)$. 
The value of $a$ can be determined experimentally and for most surfaces lies in the range 0.75-0.9 \cite{ganta2011optical, ho1970measurement}. Here, we assume a value $a=0.8$ for our calculations. The  air is assumed to be at atmospheric pressure. From Fig. \ref{fig3} it can be observed that for polymers, the heat transfer due to phonon transmission is of the order of heat conduction in air at gaps of $\approx 1$ nm. However, at such small gaps other effects which are not taken into account in the current model such as nonlocal effects, and breakdown of the simple addivity principle adopted here for calculating the van der Waals force will have to be considered for a more accurate comparison.

\section*{Acknowledgements}
This project has received funding from the European Union's Horizon 2020 research and innovation programme under the Marie Sklodowska-Curie grant agreement No 702525.



\section*{References}


\end{document}